\documentstyle[12pt]{article}

\newcommand{\be}{\begin{equation}}
\newcommand{\ee}{\end{equation}}
\newcommand{\bea}{\begin{eqnarray}}
\newcommand{\eea}{\end{eqnarray}}

\begin{document}
 \thispagestyle{empty}
 \begin{flushright}
 {MZ-TH/98-26}\\[3mm]
 {hep-th/9806248}\\[5mm]
 {June 1998}\\
\end{flushright}
 \vspace*{3cm}
 \begin{center}
 {\Large \bf
Geometrical \ approach \ to \ the \ evaluation \\[2mm]
of \ multileg \ Feynman \ diagrams\footnote{Talk presented 
at the Zeuthen Workshop ``Loops and legs in gauge theories'' 
(Rheinsberg, Germany, April 1998).
To be published in the proccedings ({\em Acta Physica Polonica}).}}
 \end{center}
 \vspace{1cm}
 \begin{center}
 A.~I.~Davydychev$^{a,b,}$\footnote{Alexander von Humboldt fellow.
On leave from the Institute for Nuclear Physics, Moscow State
University, 119899 GSP Moscow, Russia.
E-mail: davyd@thep.physik.uni-mainz.de}
 \ \ and \ \
 R.~Delbourgo$^{b,}$\footnote{E-mail: Bob.Delbourgo@utas.edu.au}\\
 \vspace{1cm}
$^{a}${\em 
 Department \ of \ Physics, \ \ \ University \ of \ Mainz, \\ 
 Staudingerweg 7, D-55099 Mainz, Germany}
\\
\vspace{.3cm}
$^{b}${\em
 Physics \ Department, \ \ \ University \ of \ Tasmania, \\
 GPO Box 252-21, Hobart, Tasmania, 7001 Australia}
\end{center}
 \hspace{3in}
\begin{abstract}
A connection between one-loop $N$-point Feynman diagrams
and certain geometrical quantities in non-Euclidean
geometry is discussed.
A geometrical way to calculate the 
corresponding Feynman integrals is considered. 
\end{abstract}
  
\newpage

\section{Introduction}

As a rule, explicit results for diagrams with several external legs
possess a rather complicated analytical structure.
This structure can be better understood
if one employs a geometrical interpretation of kinematic invariants
and other quantities. For example, the singularities of
the general three-point function can be described pictorially through a
tetrahedron constructed out of the external and internal momenta.
This method can be used to derive Landau equations defining the
positions of possible singularities \cite{Landau} (see
also in \cite{Mandelstam}) and a
similar approach can be applied to the four-point function
\cite{KSW2} too.
Another known example of using geometrical ideas
is the massless three-point function with arbitrary off-shell
external momenta (see \cite{massless,DT2}).

In this paper, we briefly describe how some geometrical
ideas can be used to calculate multileg Feynman diagrams.
In particular, we show that there is a direct transition from the
Feynman parametric representation to the geometrical description
connected with an $N$-dimensional simplex.
A more detailed discussion can be found in \cite{DD}
(see also in \cite{OW}). 

\section{A simplex related to the $N$-point function}

The scalar integral corresponding to the one-loop $N$-point
function is
\begin{equation}
\label{sun}
J^{(N)}(n; \nu_1, \ldots, \nu_N )\equiv
\int \mbox{d}^n q \;  
\prod_{i=1}^N \left[\left(p_i+q\right)^2 - m_i^2 \right]^{-\nu_i} ,
\end{equation}
where $n$ is the space-time dimension and $\nu_i$ are the   
powers of the propagators. In general, it depends on 
${\textstyle{1\over2}} N(N-1)$ momenta invariants
$k_{jl}^2$ ($j<l$), where $k_{jl} \equiv p_j-p_l$,
and $N$ masses $m_i$ corresponding to the internal propagators.  
The Feynman parametric representation for the integral (\ref{sun}) 
reads
\begin{eqnarray}
\label{fp2}
J^{(N)}\left(n; \nu_1, \ldots , \nu_N \right)
&=& \mbox{i}^{1-2\Sigma\nu_i} \pi^{n/2} \;
\Gamma\left(\sum\nu_i - {\textstyle{n\over2}} \right)
     \left[\prod\Gamma\left(\nu_i\right) \right]^{-1}
\nonumber \\
&\times&
\int_0^1 \!\ldots\! \int_0^1 \!
\prod \alpha_i^{\nu_i-1} \mbox{d}\alpha_i \;   
      \delta\left( \sum\alpha_i -1 \right) \;
\nonumber \\
&\times&
\bigg[ \sum \alpha_i^2 m_i^2 \!+\!
2\!\begin{array}{c} {} \\[-2mm] {\sum\sum} \\[-2mm] {}_{j<l}
    \end{array}
\alpha_j \alpha_l m_j m_l c_{jl} \bigg]^{n/2-\Sigma\nu_i},
\end{eqnarray}
where 
\begin{equation}
\label{def_c}
c_{jl} \equiv (m_j^2 + m_l^2 - k_{jl}^2)/(2 m_j m_l) .
\end{equation}
In the region between the corresponding two-particle pseudo-threshold,
$k_{jl}^2 \!= (m_j-m_l)^2$, and the threshold, $k_{jl}^2=(m_j+m_l)^2$,
we have $|c_{jl}|<1$, and therefore in this region they can be
understood as cosines of some angles $\tau_{jl}$,
$c_{jl} = \cos\tau_{jl}$.
At the pseudo-threshold $c_{jl}=1$ and
$\tau_{jl}=0$, whereas at the threshold 
$c_{jl}=-1$ and $\tau_{jl}=\pi$.
Note that the limits of integration in eq.~(\ref{fp2})
can be extended from $(0,1)$ to $(0,\infty)$, since the actual region
of integration is defined by the $\delta$ function.
The expressions in other regions should be understood in the sense
of analytic continuation, using (when necessary) the causal prescription
for the propagators.

Let us consider a set of $N$-dimensional  
Euclidean ``mass'' vectors whose lengths are $m_i$. Let them be directed
so that the angle between the $j$-th and the $l$-th vectors
is $\tau_{jl}$. If we denote
the corresponding unit vectors as $a_i$ (so that the ``mass'' vectors  
are $m_i a_i$), we get
$(a_j \cdot a_l)=\cos\tau_{jl}=c_{jl}$.
If we put all ``mass'' vectors
together as emanating
from a common origin, they, together with the sides connecting
their ends, will define a {\em simplex} which is
the {\em basic} one for a given Feynman diagram.
In two dimensions, the simplex is just a triangle, whereas in
three dimensions we get a tetrahedron.
It is easy to see that
the length of the side connecting the ends of the $j$-th and the $l$-th
mass vectors is $\sqrt{k_{jl}^2}$, so we shall call 
it a ``momentum'' side.
In total, the {\em basic} $N$-dimensional simplex
has ${\textstyle{1\over2}}N(N+1)$ sides,
among them $N$ mass sides (corresponding to the masses
$m_1, \ldots, m_N$) and ${\textstyle{1\over2}}N(N-1)$
momentum sides (corresponding to the momenta $k_{jl}, j<l$),
which meet at $(N+1)$ vertices.
Each vertex is a ``meeting point'' for $N$ sides.
There is one vertex where all mass sides meet, 
the {\em mass meeting point}, whereas
all other vertices are meeting points for 
$(N-1)$ momentum sides and one mass side.

The matrix $\|c\|\equiv\|c_{jl}\|$ with the components 
(\ref{def_c}) is nothing but
the Gram matrix of the vectors $a_1,\ldots,a_N$. It
is associated with many geometrical properties of the
basic simplex.
In particular, we need its determinant,
\begin{equation}
\label{d(n)}
D^{(N)} \equiv \det\|c_{jl}\| .
\end{equation}
The {\em content} (hyper-volume) of the $N$-dimensional simplex 
is given by
\begin{equation}
\label{vn}
V^{(N)} = \frac{1}{N!} \left(\prod\limits_{i=1}^N m_i \right)
\sqrt{D^{(N)}} \;\; .
\end{equation}

The number of $(N-1)$-dimensional hyperfaces is $(N+1)$.
$N$ of them correspond to the $(N-1)$-point functions, which
can be obtained from the basic $N$-point function by shrinking
one of the internal propagators in turn. The last hyperface
contains only momentum sides and can be associated with 
the massless $N$-point function. The content of this $(N-1)$-dimensional
{\em momentum} hyperface is
\begin{equation}
\label{lambdan}
\Lambda^{(N)}/(N-1)! \;\; , \hspace{10mm}
\Lambda^{(N)}=\det\|(k_{jN}\cdot k_{lN})\| .
\end{equation}

Using substitutions of variables similar to those described
in refs.~\cite{Scharf,DT2}, we can transform (\ref{fp2})
into the following form:
\begin{eqnarray}
\label{fp4}
J^{(N)}\left(n; \nu_1, \ldots , \nu_N \right)
= 2 \mbox{i}^{1-2\Sigma\nu_i} \pi^{n/2}\;
\Gamma\left(\sum\nu_i \!-\! {\textstyle{n\over2}} \right)\;
\left[\prod\Gamma\left(\nu_i\right)\right]^{-1}
\; \prod m_i^{-\nu_i}
\nonumber \\
\times
\int_0^{\infty} \!\ldots\! \int_0^{\infty}
\prod \alpha_i^{\nu_i-1} \mbox{d}\alpha_i \;
\delta\left( \alpha^T \|c\| \alpha \!-\! 1 \right)
\left( \sum\frac{\alpha_i}{m_i} \right)^{\Sigma\nu_i-n},
\end{eqnarray}  
where
\begin{equation}
\label{matrix_not}
 \alpha^T \|c\| \alpha \equiv
\sum_{j=1}^N \sum_{l=1}^N c_{jl} \alpha_j \alpha_l
=  \sum \alpha_i^2  +
2\!\begin{array}{c} {} \\[-2mm] {\sum\sum} \\[-2mm] {}_{j<l}
\end{array}
\alpha_j \alpha_l c_{jl} .
\end{equation}

Consider a special case $n=N$, $\nu_1=\ldots =\nu_N=1$.
In this case, the integrand of the parametric integral in (\ref{fp4})
is just the $\delta$ function.
The integration extends over a part of a quadratic hypersurface
defined by $\alpha^T\|c\|\alpha=1$. 
We can make a rotation to the principal axes, 
$\alpha^T\|c\|\alpha\Rightarrow \sum\lambda_i\beta_i^2$, where 
$\lambda_1\ldots\lambda_N=D^{(N)}$. Let us assume that
all $\lambda_i$ are real and positive, i.e.\
the hypersurface is an $N$-dimensional ellipsoid
(if some of the $\lambda$'s are negative, the analytic 
continuation should be used).
Now we can rescale $\beta_i=\gamma_i/\sqrt{\lambda_i}$, and
the ellipsoid becomes a hypersphere.
All we need to calculate is the content of a part of this
hypersphere which is cut out (in the space of $\gamma_i$) by the images
of the hyperfaces restricting the region where all $\alpha_i$ are
positive (in the space of $\alpha_i$). This content, 
$\Omega^{(N)}$, can be understood as the $N$-dimensional   
solid angle subtended by the above-mentioned hyperfaces.

The following statement can be proved (see in \cite{DD}):
The content of the $N$-dimensional solid angle $\Omega^{(N)}$
in the space of $\gamma_i$ is equal to that at the 
mass meeting point of the basic
$N$-dimensional simplex. Moreover, the angles between the  
corresponding hyperfaces in the space of $\gamma_i$ and those
in the basic simplex are the same. Therefore, the result 
can be expressed as 
\begin{equation}
\label{uniq5}
J^{(N)}\left(N; 1, \ldots , 1 \right)  
= \mbox{i}^{1-2N} \; \pi^{N/2} \;
\frac{\Gamma\left( N/2 \right)}
     {N!} \;\;
\frac{\Omega^{(N)}}{V^{(N)}} \; .
\end{equation}
We see that $\Omega^{(N)}$ is
indeed the only thing which is to be calculated, since $V^{(N)}$ is
known through eq.~(\ref{vn}). 

Moreover, $\Omega^{(N)}$ is nothing
but the content of a {\em non-Euclidean} $(N-1)$-dimensional simplex
calculated in the spherical (or hyperbolic,
depending on the signature of the eigenvalues $\lambda_i$) space
of constant curvature.
The sides of this non-Euclidean simplex are equal to
the angles $\tau_{jl}$. Therefore, the problem of calculating  
Feynman integrals is intimately connected with the problem of
calculating the content of a simplex in non-Euclidean geometry.

In the general case, when $\Sigma\nu_i\neq n$, we need some
modification of the above transformations (see ref.~\cite{DD}). 
In particular,
when $\nu_1=\ldots =\nu_N=1$ (but $N\neq n$) the result
generalizing eq.~(\ref{uniq5}) reads 
\begin{equation}
\label{gen3}
J^{(N)}(n; 1, \ldots, 1) = \mbox{i}^{1-2N} \pi^{n/2} \;
\Gamma\left(N-{\textstyle{n\over2}}\right) \;
\frac{m_0^{n-N} \; \Omega^{(N;n)}}{N! \; V^{(N)}} ,
\end{equation}
with
\begin{equation}
\label{omegann}
\Omega^{(N;n)} \equiv
\begin{array}{c} {} \\ \int\limits \ldots \int\limits \\   
                 {}^{\Omega^{(N)}} \end{array}
\frac{\mbox{d}\Omega_N}{\cos^{n-N}\theta} .
\end{equation}
Geometrically, $\theta$ can be understood as the angle between 
the ``running'' vector of integration and the direction 
of the height of the basic simplex, $H_0$. 
Denoting the angle  between $H_0$ and the $i$-th
mass side as $\tau_{0i}$, we get 
\begin{equation}
\label{m0}
\cos\tau_{0i}=m_0/m_i ,
\hspace{8mm}
m_0\equiv |H_0|=\left(\prod_{i=1}^N m_i\right)
\sqrt{D^{(N)}/\Lambda^{(N)}},
\end{equation}
with $\Lambda^{(N)}$ defined by eq.~(\ref{lambdan}).

Furthermore, we can use the height $H_0$ to split the basic
$N$-dimensional simplex into $N$ rectangular ones, each time
replacing one of the mass sides, $m_i$, by $H_0$ ($|H_0|=m_0$).
In this way, we split $\Omega^{(N)}$ into $N$ parts 
$\Omega_i^{(N)}$. Therefore, the Feynman integral (\ref{gen3})
can be presented as
\begin{equation}
\label{split1_jn}
J^{(N)}(n;1, \ldots, 1) = \sum_{i=1}^N \frac{V_i^{(N)}}{V^{(N)}} \;
J_i^{(N)}(n;1, \ldots, 1) ,
\end{equation}
where $J_i^{(N)}$ denotes the integral associated with the
$i$-th rectangular simplex, whilst $V_i^{(N)}$ is the known
content of this simplex. 

\section{Some examples}

For the two-point function, the basic simplex is a triangle with
the sides $m_1$, $m_2$ and $\sqrt{k_{12}^2}$.
Furthermore, $V^{(2)}={\textstyle{1\over2}}m_1m_2\sin\tau_{12}$,
$\Omega^{(2)}=\tau_{12}$ and $\Lambda^{(2)}=k_{12}^2$.
In two dimensions, from (\ref{uniq5}) we obtain the well-known result
\begin{equation}
J^{(2)}(2;1,1) = \frac{\mbox{i}\pi}{m_1 m_2} \;
                 \frac{\tau_{12}}{\sin\tau_{12}} ,
\end{equation}
In four dimensions, introducing dimensional
regularization \cite{dimreg}, we get
\begin{equation}
\label{J^2(4,1,1)}
J^{(2)}(4-2\varepsilon;1,1)
=\mbox{i}\pi^{2-\varepsilon}\Gamma(\varepsilon)
\frac{m_0^{1-2\varepsilon}}{\sqrt{\Lambda^{(2)}}} \;
\left\{ \Omega_1^{(2; 4-2\varepsilon)}
+ \Omega_2^{(2; 4-2\varepsilon)} \right\} ,
\end{equation}
with (see, e.g., in \cite{2pt})
\begin{equation}
\Omega_i^{(2; 4-2\varepsilon)} =
\int_0^{\tau_{0i}}
\frac{\mbox{d}\theta}{\cos^{2-2\varepsilon}\theta}
= 2 \tan\tau_{0i} \;\;
_2F_1\left( \left.
\begin{array}{c} 1/2, \; \varepsilon \\ 3/2 \end{array}
\right| -\tan^2\tau_{0i} \right) ,
\end{equation}  
where $\tau_{01}$ and $\tau_{02}$ are defined in eq.~(\ref{m0}),
$\tau_{01}+\tau_{02}=\tau_{12}$.

For the three-point function,
the three-dimensional basic simplex is a tetrahedron with three
mass sides (the angles between these mass sides are $\tau_{12},
\tau_{13}$ and $\tau_{23}$) and three momentum sides.
The volume of this tetrahedron is defined by eq.~(\ref{vn})
at $N=3$.
Furthermore, $\Omega^{(3)}$ is the usual solid angle at the vertex
derived by the mass sides. Its value can be defined as the area of
a part of the unit sphere cut out by the three planar faces adjacent
to the vertex; in other words, this is the area of a spherical triangle
corresponding to this section. The sides of this spherical triangle
are obviously equal to the angles $\tau_{12}, \tau_{13}$ and $\tau_{23}$
while its angles, $\psi_{12}, \psi_{13}$ and $\psi_{23}$, are equal
to those between the plane faces.
The area of this spherical triangle is
\begin{equation}
\label{solid_angle}
\Omega^{(3)}=\psi_{12}+\psi_{13}+\psi_{23}-\pi 
= 2 \arctan
\left(\sqrt{D^{(3)}}/(1\!+\!c_{12}\!+\!c_{13}\!+\!c_{23})\right).
\end{equation}
Finally, the result
\begin{equation}
\label{J(3)}
J^{(3)}(3;1,1,1) = -\frac{\mbox{i} \pi^2}{2 m_1 m_2 m_3} \;
                    \frac{\Omega^{(3)}}{\sqrt{D^{(3)}}} 
\end{equation}
corresponds to one
obtained in \cite{Nickel} in a different way.

If we consider the four-dimensional three-point function, the
only (but very essential!) difference is that we should divide the
integrand by $\cos\theta$. We split the spherical triangle
with the sides $\tau_{12}, \tau_{13}$ and $\tau_{23}$ into
three spherical triangles, corresponding to the solid angles
of rectangular tetrahedra. Calculating the corresponding 
integrals, we obtain the result in terms of the dilogarithms,
or the Clausen function (see e.g. in \cite{'tHV-79}).

For the four-point function, 
the corresponding four-dimensional simplex has four mass sides
and six momentum sides.
It has five vertices and five three-dimensional
hyperfaces. Four of these hyperfaces are the {\em reduced} ones,
corresponding to three-point functions, whereas
the fifth one is the momentum hyperface.
This four-dimensional simplex is completely defined by
its mass sides
$m_1, m_2, m_3, m_4$ and six ``planar'' angles between them,
$\tau_{12}, \tau_{13}, \tau_{14}, \tau_{23}, \tau_{24}$ and $\tau_{34}$.
The content (hyper-volume) of this simplex is given by
eq.~(\ref{vn}) at $N=4$, with $D^{(4)}= \det\|c_{jl}\|$.

The four-dimensional four-point function can be 
exhibited as (cf.\ eq.~(\ref{uniq5}))
\begin{equation}
\label{J4}
J^{(4)}(4;1,1,1,1)= {\textstyle{1\over12}}\; \mbox{i} \pi^2 \;
\frac{\Omega^{(4)}}{V^{(4)}}
= \frac{2 \; \mbox{i} \pi^2}{ m_1 m_2 m_3 m_4 } \;
\frac{\Omega^{(4)}}{\sqrt{D^{(4)}}} .
\end{equation}
So, the main problem is how to calculate $\Omega^{(4)}$.

In four dimensions, $\Omega^{(4)}$ is the value of the
four-dimensional generalization of the solid   
angle at the mass meeting point of the simplex.
In the spherical case, it can be defined as the volume of a part
of the unit hypersphere
which is cut out from it by the four three-dimensional
reduced hyperfaces, each hyperface involving three mass sides
of the simplex.
This hyper-section is a three-dimensional spherical tetrahedron 
whose six sides (edges) are equal to the angles $\tau_{jl}$.
In the hyperbolic case, this is a hyperbolic
tetrahedron whose volume can be obtained by analytic
continuation.

Unfortunately, there are no {\em simple}
relations like (\ref{solid_angle}) which might make it possible to  
express the volume of a spherical (or hyperbolic) tetrahedron
in terms of its sides or  
dihedral angles. In fact, calculation of this volume in an elliptic
or hyperbolic space is a well-known problem of non-Euclidean
geometry (see e.g. in \cite{Lobachevsky}).
A standard way to solve this problem, say in spherical space,
is to split an arbitrary tetrahedron into a set of birectangular ones.
The volume of a birectangular tetrahedron is known and 
can be expressed in terms of Lobachevsky or Schl\"afli functions   
which can be related to dilogarithms or Clausen function
(see in \cite{Coxeter}).
Different ways of splitting the non-Euclidean tetrahedron
can be used to reduce the number of dilogarithms (or related
functions) involved (cf.\ in \cite{'tHV-79,DNS}). 

\section{Conclusion}

We have shown that there is a direct link between Feynman 
parametric representation of a one-loop $N$-point function
and the basic simplex in $N$-dimensional Euclidean space.
In the case $N=n$ (where $n$ is the space-time dimension), 
the result for the Feynman integral turns out to
be proportional to the ratio of an $N$-dimensional solid angle
at the meeting point of the mass sides to the content of
the $N$-dimensional basic simplex.
For the four-dimensional four-point
function, the representation (\ref{fp4})  provides a very 
interesting connection with the volume
of the non-Euclidean (spherical or hyperbolic) tetrahedron.

In the general case ($N\neq n$), the height of the basic simplex, 
$H_0$, plays an essential role in calculation of the integrals.
It is used to split the basic Euclidean simplex into $N$ rectangular
simplices.
When $N<n$, this splitting simplifies the calculation of
separate integrals.
When $N=n+1$, each integral $J_i^{(N)}$ (see eq.~(\ref{split1_jn}))
corresponding to one
of the resulting rectangular tetrahedra can be reduced
to an $(N-1)$-point function (cf.\ also in \cite{reduction,Nickel}).

In the resulting expressions, all arguments of functions arising
possess a straightforward geometrical meaning in terms of the
dihedral angles, etc.
In particular, this is quite useful for
choosing the most convenient kinematic variables to describe the
$N$-point diagrams. We suggest that this approach can help in 
understanding the geometrical structure of
loop integrals with several external legs, as well as the
structure of phase-space integrals.
We also note a connection with 3-loop vacuum
graphs in three dimensions \cite{Br}.

\vspace{3mm}

{\bf Acknowledgements}.
A.D. is grafeful for hospitality to the Physics Department, 
University of Tasmania, where this work was started. 
A.D.'s research was essentially supported
by the Alexander von Humboldt Foundation, and partly
by the grants INTAS--93-0744, RFBR--98-02-16981
and Volkswagen--I/73611.
A.D. is grateful to the organizers (DESY-Zeuthen) for partial
support of his participation in the Rheinsberg workshop.


\begin{thebibliography}{99}

\bibitem{Landau}
L.D.~Landau, {\em Nucl.~Phys.} {\bf 13}, 181 (1959);\\
G.~K\"allen, A.~Wightman,
{\em Mat.~Fys.~Skr.~Dan.~Vid.~Selsk.} {\bf 1} (No.6), 1 (1958).

\bibitem{Mandelstam} 
S.~Mandelstam, {\em Phys.~Rev.} {\bf 115}, 1742 (1959); \\
R.E.~Cutkosky, {\em J.~Math.~Phys.} {\bf 1}, 429 (1960); \\
R.J.~Eden, P.V.~Landshoff, D.I.~Olive, J.C.~Polkinghorne,
{\em The analytic $S$-matrix} (Cambridge UP, 1966).

\bibitem{KSW2} R.~Karplus, C.M.~Sommerfield, E.H.~Wichmann,
             {\em Phys.~Rev.} {\bf 114}, 376 (1959);\\
A.C.T.~Wu, {\em Mat.~Fys.~Medd.~Dan.~Vid.~Selsk.}
             {\bf 33} (No.3), 1 (1961).
   
\bibitem{massless}
J.S.~Ball, T.-W.~Chiu, {\em Phys.~Rev.} {\bf D22}, 2550 (1980); \\
A.I.~Davydychev, {\em J.~Phys.} {\bf A25}, 5587 (1992); \\
H.J.~Lu, C.A.~Perez, preprint SLAC-PUB-5809 (1992).

\bibitem{DT2}
A.I.~Davydychev, J.B.~Tausk, {\em Phys.~Rev.} {\bf D53}, 7381 (1996).

\bibitem{DD} A.I.~Davydychev, R.~Delbourgo,
{\em J.~Math.~Phys.}, 
to appear, hep-th/9709216.

\bibitem{OW} N.~Ortner, P.~Wagner,
{\em Ann.~Inst.~Henri~Poincar\'e (Phys.~th\'eor.)} {\bf 63}, 81 (1995).

\bibitem{Scharf} R.~Scharf, Doctoral Thesis, W\"urzburg (1994); \\ 
R.~Scharf, J.B.~Tausk, {\em Nucl.~Phys.} {\bf B412}, 523 (1994).

\bibitem{dimreg}
G.~'t~Hooft, M.~Veltman, {\em Nucl.~Phys.} {\bf B44}, 189 (1972);\\
C.G.~Bollini, J.J.~Giambiagi, {\em Nuovo~Cimento} {\bf 12B}, 20 (1972).

\bibitem{2pt}
U.~Nierste, D.~M\"uller, M.~B\"ohm, {\em Z.~Phys.} {\bf C57}, 605 (1993);\\
F.A.~Berends, A.I.~Davydychev, V.A.~Smirnov,
                  {\em Nucl.~Phys.} {\bf B478}, 59 (1996);\\
E.~Remiddi, {\em Nuovo Cimento} {\bf 110A}, 1435 (1997). 

\bibitem{Nickel} B.G.~Nickel, {\em J.~Math.~Phys.} {\bf 19}, 542 (1978).

\bibitem{'tHV-79} G.~'tHooft and M.~Veltman, {\em Nucl.~Phys.}
             {\bf B153}, 365 (1979); \\
A.~Denner, {\em Fortschr.~Phys.} {\bf 41}, 307 (1993).

\bibitem{DNS} A.~Denner, U.~Nierste, R.~Scharf, {\em Nucl.~Phys.}
             {\bf B367}, 637 (1991).

\bibitem{Lobachevsky} N.I.~Lobatschefsky, {\em Imagin\"are Geometrie},
           Kasaner Gelehrte Schriften, 1836 (\"Uber\-setzung mit
           Anmerkungen von H.~Liebmann, Leipzig, 1904);\\
L.~Schl\"afli, {\em Quart.~J.~Math.} {\bf 3}, 54 (1860); 
     {\bf 3}, 97 (1860); \\
     {\em Gesammelte matematische Abhandlungen}, Band II
     (Birkh\"auser, Basel, 1953).
     
\bibitem{Coxeter}
H.S.M.~Coxeter, {\em Quart.~J.~Math.} {\bf 6}, 13 (1935);\\
E.B.~Vinberg, {\em Uspekhi~Mat.~Nauk} {\bf 48} (No.2), 17 (1993)
        [{\em Russian~Math.~Surveys} {\bf 48} (No.2), 15 (1993)];\\
R.~Kellerhals, in {\em Structural properties
of Polylogarithms} (ed. L.~Lewin), AMS Math. Surveys and Monographs,
vol.~37, p.301 (1991).

\bibitem{reduction}
L.M.~Brown, {\em Nuovo~Cimento} {\bf 22}, 178 (1961);\\
F.R.~Halpern, {\em Phys.~Rev.~Lett.} {\bf 10}, 310 (1963);\\
G.~K\"all\'en and J.~Toll, {\em J.~Math.~Phys.} {\bf 6}, 299 (1965);\\
B.~Petersson, {\em J.~Math.~Phys.} {\bf 6}, 1955 (1965);\\
D.B.~Melrose, {\em Nuovo~Cimento} {\bf 40A}, 181 (1965);\\
W.L.~van~Neerven, J.A.M.~Vermaseren, {\em Phys.~Lett.}
     {\bf B137}, 241 (1984);\\
Z.~Bern, L.~Dixon, D.A.~Kosower, {\em Phys.~Lett.} {\bf B302}, 299 (1993);
     {\em Nucl.~Phys.} {\bf B412}, 751 (1994).

\bibitem{Br} D.J.~Broadhurst, preprint~OUT-4102-74 (1998), 
hep-th/9806174.

\end{thebibliography}
\end{document}